 \documentclass[final,3p,times,twocolumn]{elsarticle}

\usepackage{epsfig,bm}
\usepackage{graphics}
\usepackage{amsmath}
\usepackage{mathrsfs}
\usepackage[normalem]{ulem}  
\usepackage[dvips]{color} 

\renewcommand\sout{\bgroup \color{red} \ULdepth=-.5ex \ULset}

\biboptions{sort&compress}

\begin{document}

\begin{frontmatter}


\title{Hot medium effects on $J/\psi$ production in $p+\textrm{Pb}$ collisions at $\sqrt{s_{\rm NN}}=5.02$\ TeV}

\author[ic]{Yunpeng Liu}
\author[ic,ph]{Che Ming Ko}
\author[ic]{Taesoo Song}

\address[ic]{Cyclotron Institute, Texas A$\&$M University, College Station, Texas 77843, USA}
\address[ph]{Department of Physics and Astronomy, Texas A$\&$M University, College Station, Texas 77843, USA}%
\begin{abstract}
   Based on a kinetic description of $J/\psi$ dissociation and production in  an expanding quark-gluon plasma  that is described by a 2+1 dimensional ideal hydrodynamics, we have studied the hot medium effects on $J/\psi$ production in $p+\textrm{Pb}$ collisions at $\sqrt{s_{\rm NN}}=5.02$\ TeV.  Including also the cold nuclear matter effects, we are able to reproduce recent experimental results  on the nuclear modification factor $R_{p\textrm{Pb}}(J/\psi)$ measured by the ALICE Collaboration.  We have also made predictions for the  $R_{p\textrm{Pb}}$ of $J/\psi$ and the double ratio $R_{p\textrm{Pb}}^{\textrm{pro}}(\psi')/R_{p\textrm{Pb}}^{\textrm{pro}}(J/\psi)$ of prompt quarkonia produced in the most central 10\%  $p+\textrm{Pb}$ collisions.  We find that different from the cold nuclear matter effects, the $R_{p\textrm{Pb}}(J/\psi)$ is slightly smaller than that in the minimum bias collisions, and the double ratio is  significantly less  than  one at backward rapidity.
\end{abstract}


\end{frontmatter}



\section{Introduction}
One of the important probes of the  quark-gluon plasma (QGP) produced in relativistic heavy ion collisions is the  suppressed production of $J/\psi$  as a result of  the color screening of the potential between charm and anticharm quarks, as first suggested in Ref.~\cite{Matsui:1986dk}. Studies of $J/\psi$ production  in $p+p$, $p+\textrm{A}$, $d+\textrm{A}$, and $\textrm{A}+\textrm{A}$ collisions~\cite{Gonin:1996wn, Abreu:2000ni, Adare:2006kf, Adler:2005ph, Adare:2011yf, Adamczyk:2012ey, Manceau:2013zta, Chatrchyan:2012np, Abelev:2012rv} have indicated, however, that besides this hot  medium effect~\cite{Matsui:1986dk, Gonin:1996wn}, there is also the normal suppression~\cite{Abreu:1999nn}  due to cold nuclear matter effects. The cold nuclear matter effects~\cite{Gerschel:1988wn, Gousset:1996xt, McLerran:1998nk, Kopeliovich:1991pu, Kopeliovich:1993pw, Kopeliovich:2003tz, Iancu:2003xm, Ferreiro:2008wc} include the nuclear absorption~\cite{Vogt:1999cu} that the initially produced $J/\psi$ can be destroyed by the passing nucleons, the Cronin effect~\cite{Cronin:1974zm} that the gluons gain transverse momentum before fusing into a $J/\psi$ due to the scattering with other nucleons, and the shadowing effect~\cite{Li:2001xa} that the parton distribution function in a nucleus is different from that in free nucleons.  Furthermore, there are other hot  medium effects due to the gluon dissociation of $J/\psi$ that have been described in terms of various theoretical models~\cite{Karsch:2005nk, Brambilla:1999xf, Peskin:1979va, Grandchamp:2001pf}.  Moreover, the charm and anticharm quarks in the  produced QGP may combine  to form $J/\psi$s. This regeneration process has been studied either using the statistical hadronization model~\cite{BraunMunzinger:2000px} or included in the kinetic approach as the inverse  of the dissociation process~\cite{Thews:2000rj, Rapp:2003vj, Yan:2006ve}. The regeneration effect is expected to be large in $\textrm{A}+\textrm{A}$ collisions at the Large Hadron Collider (LHC)~\cite{Liu:2009nb, Zhao:2011cv} due to the  appreciable number of charm quarks and anticharm quarks produced in these collisions. Its contribution to $J/\psi$ production in heavy ion collisions  at the Super Proton Synchrotron (SPS) can also be important due to the canonical enhancement effect in spite of the  small number of charm and anticharm quarks in collisions at lower energies~\cite{Gorenstein:2000ck, Grandchamp:2002wp}. Also, the study of $J/\psi$\ has  been extended to its excited states~\cite{Santos:2004zz, Moon:2012fr, Scomparin:2012vq}, bottomonia~\cite{Reed:2011zza, Grandchamp:2005yw, Song:2011nu} and $B_c$\ mesons~\cite{Schroedter:2000ek, Liu:2012tn}.

For  cold nuclear matter effects, it has been suggested that they can be understood from studying $p+\textrm{A}$ or $d+\textrm{A}$ collisions because  a hot  medium is not expected to be formed in these collisions. This has been the case at the SPS and the Relativistic Heavy Ion Collider (RHIC)~\cite{Alessandro:2006jt, Alessandro:2004ap, Adams:2003qm, Adler:2005ph, Adare:2007gn, Adler:2006xd, Adams:2006nd}.  At the LHC, because of the higher multiplicity of produced particles, especially in central collisions, final-state hot  medium effects can also be present besides the initial-state cold nuclear matter effects. For example, the multiplicity of charged hadrons $dN_{ch}/dy$ in $p+\textrm{Pb}$ collisions at $\sqrt{s_{\textrm{NN}}}=5.02$ TeV~\cite{ALICE:2012xs} is around $50$ in the most central 5\% collisions~\cite{Albacete:2013ei}. Since the energy carried by these charged particles and other neutral particles is initially localized in a volume of the size of a proton, the energy density at the beginning of the collision can be sufficient high for the formation of  a QGP. 

Recently, the ALICE Collaboration has measured the nuclear modification factor of $J/\psi$\ in $p+\textrm{Pb}$ collisions at $\sqrt{s_{\rm NN}}=5.02$\ TeV~\cite{Manceau:2013zta}. Although the cold  nuclear matter effects  have been studied~\cite{Albacete:2013ei}, there is not yet any study on the hot  medium effects in these collisions.  In this paper, we  use a transport model~\cite{Liu:2009nb, Liu:2012zw}, which was previously used to  study $J/\psi$ suppression and regeneration at RHIC and LHC, to  study the nuclear modification factor and the  the double ratio $R_{p\textrm{Pb}}^{\textrm{pro}}(\psi')/R_{p\textrm{Pb}}^{\textrm{pro}}(J/\psi)$ of prompt quarkonia produced in $p+\textrm{Pb}$ collisions.

This paper is organized as follows. In Sec.~\ref{se_bulk}, we give a short description of the 2+1 dimensional hydrodynamic model used in the present study for describing the evolution dynamics of  produced QGP. The transport approach used for studying the dissociation and regeneration of $J/\psi$ in the QGP is then discussed in Sec.~\ref{se_transport}. Results from the present study are shown in Sec.~\ref{se_results}. Finally, a short conclusion is given in Sec.~\ref{se_conclusions}.

\section{The bulk dynamics}\label{se_bulk}

We describe the bulk dynamics of $p+\textrm{Pb}$ collisions at $\sqrt{s_{\textrm{NN}}}=5.02$ TeV by the 2+1 dimensional ideal hydrodynamic equation 
\begin{eqnarray}
   \partial_{\mu}T^{\mu\nu}&=& 0
\end{eqnarray}
with the boost invariant condition $v_z=z/t$, where $T^{\mu\nu}=(\epsilon+p)u^{\mu}u^{\nu}-g^{\mu\nu}p$\ is the energy-momentum tensor in terms of the energy density $\epsilon$, pressure $p$, and four velocity $u$.  For the equation of state $\epsilon(p)$, we use that of Ref.~\cite{Liu:2009nb} based on an ideal gas of  massless partons for the QGP and massive  hadrons for the hadronic matter, together with a bag constant $B=(236\textrm{ MeV})^4$  that leads to a critical temperature $T_c=165$ MeV~\cite{Aoki:2006br, Borsanyi:2010bp, Aoki:2006we}. Since most colliding nucleons pass through each other without much stopping, we take the net baryon density of produced matter to be zero. 

\begin{figure}[!hbt]
    \centering
    \includegraphics[width=0.48\textwidth]{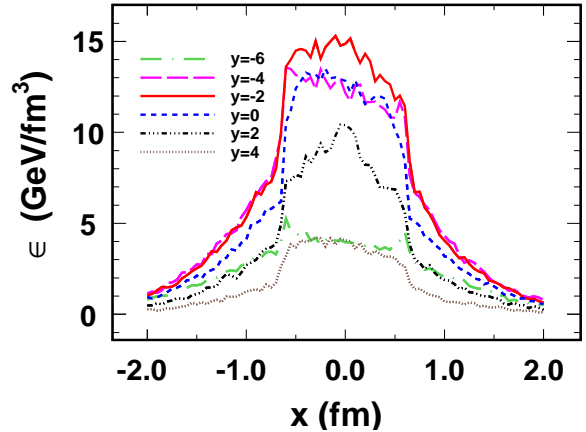}
    \caption{(Color online) Initial energy density $\epsilon$ at proper time $\tau=0.6$ fm$/c$ in the most central $10\%$  $p+\textrm{Pb}$ collisions at $\sqrt{s_{\rm NN}}=5.02$ TeV from the AMPT model as a function of the coordinate $x$\ relative to the center of the proton along the direction of the impact parameter $b$  for different rapidity $y$.}
    \label{fig_epsilon}
\end{figure}

For the initial conditions of the hydrodynamic evolution, they are  generated from a multiple phase transport (AMPT) model~\cite{Lin:2004en} (version: v2.26t1).  Specifically,  we run $9\times 10^5$ $p+\textrm{Pb}$ events  and select the most  central 10\% collisions to obtain the energy density $\epsilon$.  With the centrality  determined by the multiplicity of initially produced partons in the whole rapidity range, which is approximately equal to that determined by the multiplicity of final charged hadrons, the corresponding average impact parameter is  $\langle b \rangle=3.3$ fm.  By shifting the origin of the transverse coordinates to the center of the proton in the transverse plane before taking an average over all selected events and with the center of the Pb nucleus lying on the negative $x$-axis, the resulting energy density at an initial proper time $\tau=0.6$ fm/$c$ as a function of coordinate $x$  is shown in Fig.~\ref{fig_epsilon}  for different rapidity $y$ with the
 proton  and Pb rapidities taken to be positive and  negative, respectively. The maximum energy density $\epsilon_{\rm max}\approx 15$\ GeV/fm$^3$ appears at rapidity $y=-2$ and the corresponding temperature $T_{\rm max}$  calculated from the noninteracting QGP model is about $290$\ MeV, which is between the maximum  values in $\textrm{A}+\textrm{A}$\ collisions at SPS and RHIC.  For the centrality bin of $40\textrm{-}60\%$, in which the multiplicity of charged hadrons is very close to that in the minimum bias case~\cite{Albacete:2013ei}, the  initial energy density $\epsilon$  is  about a factor of $3$ smaller than that in the most  central 10\% bin.

\begin{figure}[!hbt]
    \centering
    \includegraphics[width=0.48\textwidth]{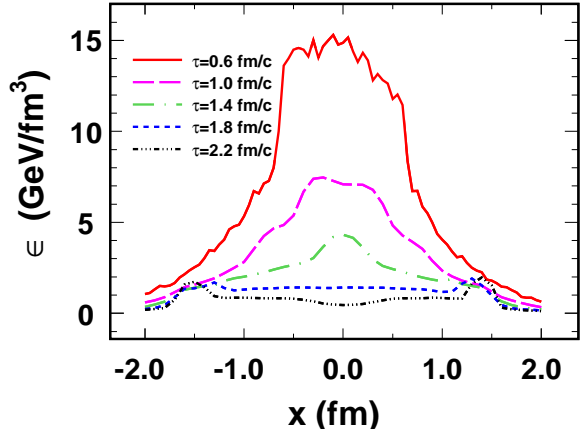}
    \caption{(Color online) Energy density $\epsilon$ at rapidity $y=-2$ in the most central $10\%$  $p+\textrm{Pb}$ collisions at $\sqrt{s_{\rm NN}}=5.02$ TeV  as a function of the coordinate $x$\ relative to the center of the proton along the direction of the impact parameter $b$ at different proper time $\tau$.}
    \label{fig_epsilon_tau}
\end{figure}

Figure~\ref{fig_epsilon_tau} shows the energy density at rapidity $y=-2$  for different proper time $\tau$  during 
hydrodynamic evolution. It  is seen that  the energy density at the center decreases quickly after 
$\tau_0=0.6$ fm$/c$ proper time. For example,  at $\tau=1.8$ fm$/c$  the energy density at the center is  
reduced by one order of magnitude from its value at the initial time $\tau_0=0.6$ fm$/c$. 
This is due to the fast expansion of the produced matter not only in the longitudinal direction but also 
in the transverse direction that is 
driven by the large pressure or energy density gradient resulting from the high initial energy 
density  and the smaller size of  the 
matter.  Indeed, the transverse diameter of the produced matter increases by about  $2$ fm from $\tau_0=0.6$ fm$/c$ to $\tau=1.8$ fm$/c$, indicating a transverse expansion velocity close to half the velocity of light. As a result, the energy density at the center decreases quickly and becomes even smaller than that in the peripheral region at later times.

\section{The Transport approach}\label{se_transport}

Because of color screening in QGP, a $J/\psi$ cannot survive above  a dissociation temperature $T_D$. We take the velocity dependence of $T_D$ from Ref.~\cite{Liu:2012zw}  by including the effect due to the motion of heavy charm and anticharm quarks in the QGP on the medium response, which leads to a higher dissociation temperature for a faster moving $J/\psi$.   With the surviving $J/\psi$ described by a distribution function $f({\bf x}, {\bf p}, t)$ in  the phase space  of coordinate ${\bf x}$ and momentum ${\bf p}$, its time dependence then satisfies  the  transport equation
\begin{eqnarray}
   \partial_t f+{\bf v}\cdot\nabla f& =& -\alpha f + \beta,
   \label{eq_transport}
\end{eqnarray}
 where the left hand side are kinetic terms  describing the free streaming of $J/\psi$, which is responsible  for the leakage effect~\cite{Zhuang:2003fu}. On the right hand side, $\alpha$\ and $\beta$ are the dissociation rate and the regeneration rate, respectively.  For the dissociation rate, it is 
given by
\begin{eqnarray}
   \alpha(T,u,p)&=& \frac{N_g}{E}\int\frac{d{\bf k}}{(2\pi)^3E_g}{k}\cdot{p}f_g(T,u, k)\sigma_D.\nonumber\\
   \label{eq_alpha}
\end{eqnarray}
In the above, $p=(E, {\bf p})$\ and $k=(E_g, {\bf k})$\ are the 4-momentum of $J/\psi$\ and gluon, respectively; $N_g=16$ and $f_g$ are the  
degeneracy and thermal distribution of gluons, which depends on the local temperature $T$ and velocity $u$; and $\sigma_D$\ is the cross section  for the dissociation process $J/\psi+g\rightarrow c+\bar{c}$ and is taken to be that calculated  in Refs.~\cite{Liu:2009nb, Manceau:2013zta} based on the operator expansion method of Ref.~\cite{Peskin:1979va}.  

The regeneration rate due to the inverse process $c+\bar{c}\rightarrow J/\psi + g$ is given by
\begin{eqnarray}
   &&\beta(T,u,p)\nonumber\\
   &=& \frac{C_{ce}}{2E}\int\frac{d{\bf k}}{(2\pi)^32E_g}\frac{d{\bf q}_c}{(2\pi)^32E_c}\frac{d{\bf q}_{\bar c}}{(2\pi)^32E_{\bar c}}W(s)\nonumber\\
   &\times&rf_cf_{\bar c}(1+f_g)(2\pi)^4\delta^{(4)}(p+k-q_c-q_{\bar c}).
   \label{eq_beta}
\end{eqnarray}
 In the above, $p$\ and $k$\ have the same meaning as in Eq.~(\ref{eq_alpha}), and $q_c=(E_c, {\bf q}_c)$\ and $q_{\bar c}=(E_{\bar c}, {\bf q}_{\bar c})$\ are the momenta of charm and anticharm quarks. The transition rate $W$\ is related to the $J/\psi$ dissociation cross section $\sigma_D$\ by the detailed balance relation.  For the distribution function $f_c$ of charm quarks,  it is taken to have following factorized form for simplicity: 
\begin{eqnarray}
   f_c({\bf x}, {\bf p})&=& \rho_c({\bf x})\frac{N}{e^{q_c\cdot u/T}+1},
\end{eqnarray}
where $\rho_c$\ is the number density of charm quarks and satisfies the conservation equation $\partial_{t}\rho_c+\nabla\cdot(\rho_c{\bf v})=0$ with ${\bf v}$\ being the velocity of the medium, and $N\equiv \left[(2\pi)^{-3}\int d {\bf q}_c\left(e^{q_c\cdot u/T}+1\right)^{-1}\right]^{-1}$\ is the normalization factor  for the Fermi distribution.  For the anticharm quarks, their distribution function  $f_{\bar c}$  has a similar form. 

The factors $r$ and $C_{ce}$ in Eq.(\ref{eq_beta}) take into account, respectively, the charm quark non-equilibrium effect and the canonical effect due to few pairs of charm and anticharm quarks produced in a $p+\textrm{Pb}$ collision, that are neglected in Ref.~\cite{Liu:2009nb, Liu:2012zw}.  Because of the short QGP lifetime, charm quarks are not expected to be fully thermalized~\cite{Grandchamp:2002wp, Zhao:2007hh, Song:2012at},  leading to a reduction of the regeneration rate through the relaxation factor $r=(1-\exp(-\tau/\tau_r))$  introduced in Ref.~\cite{Grandchamp:2002wp}. For  the relaxation time $\tau_r$, we take its value to be  $7$ fm$/c$ as that used in Ref.~\cite{Zhao:2007hh} for $\textrm{Pb}+\textrm{Pb}$ collisions at SPS energy and $\textrm{Au}+\textrm{Au}$ collisions at RHIC energy.   The canonical effect has been shown to be responsible for the suppressed production of strange mesons~\cite{Rafelski:1980gk, Ko:2000vp} when few strange quarks are produced. With few pairs of charm and anticharm quarks produced in an event, the same effect becomes important and enhances charmonia production in heavy ion collisions~\cite{Gorenstein:2000ck, Grandchamp:2003uw}. It modifies the relation between the number of directly produced charm quarks $N_c^{\rm dir}$ and the thermally equilibrated $D$-meson and $J/\psi$ numbers $N_D^{\rm th}$ and $N_{J/\psi}^{\rm th}$~\cite{Kostyuk:2005zd, Liu:2012tn} to
\begin{eqnarray}
   N_{c}^{\rm dir}&=& \gamma N_D^{\rm th}+(1+1/N_c^{\rm dir})\gamma^2N_{J/\psi}^{\rm th},
\end{eqnarray}
where $\gamma$ is the fugacity of the charm quark, resulting in an enhanced production of $J/\psi$ by the  factor $(1+1/N_c^{\rm dir})$.  Since a charm and anticharm quark pair is produced at same rapidity, the canonical enhancement factor for $J/\psi$ production in a rapidity bin $\Delta y=1$ becomes  $C_{ce}=1+1/(dN_c^{\rm dir}/dy)$~\cite{Andronic:2006ky, Kostyuk:2005zd, Liu:2012tn}. This canonical effect can be included in a more accurate way through kinetic approaches, which will be studied in the future.

For the initial distribution of $J/\psi$, which is needed for solving the transport equation, it is obtained from the  Glauber model with the inclusion of initial-state  cold nuclear matter effects. Since there is no empirical $J/\psi$ production  cross section  $d\sigma_{pp}/dy$ for $p+p$ collisions at $\sqrt{s_{\textrm{NN}}}=5.02$ TeV, its value is obtained  from the experimental results at lower ($\sqrt{s_{\textrm{NN}}}=2.76$ TeV)~\cite{Abelev:2012kr} and higher ($\sqrt{s_{\textrm{NN}}}=7$ TeV)~\cite{Aamodt:2011gj} energies by interpolation, and the result is $d\sigma_{pp}/dy=5.68$ $\mu$b at midrapidity. Similarly,  the first two moments of the $J/\psi$ transverse momentum distribution in the forward rapidity region $2.5<y<4$ are  found to be $\langle p_T\rangle = 2.37$\ GeV and $\langle p_T^2 \rangle = 7.73$\ GeV$^2$.  

Assuming a power-law distribution for the $J/\psi$ transverse momentum at a given rapidity~\cite{Bossu:2011qe}
\begin{eqnarray}\label{dydpt}
   \frac{d^2\sigma}{dydp_T}&=& \frac{d\sigma}{dy}\frac{2(n-1)}{D_y}p_T\left(1+\frac{p_T^2}{D_y}\right)^{-n},
\end{eqnarray}
 we obtain the ratio
\begin{eqnarray}
   \frac{\langle p_T \rangle^2_y}{\langle p_T^2\rangle_y}&=& (n-2)a_n^2
\end{eqnarray}
with $a_n=(n-1)B(n-3/2,3/2)$ and $B$  being the Beta function. We then find $n=5.05$ from the above 
values for $\langle p_T\rangle$ and $\langle p_T^2\rangle$. 

\begin{figure}
    \centering
    \includegraphics[width=0.48\textwidth]{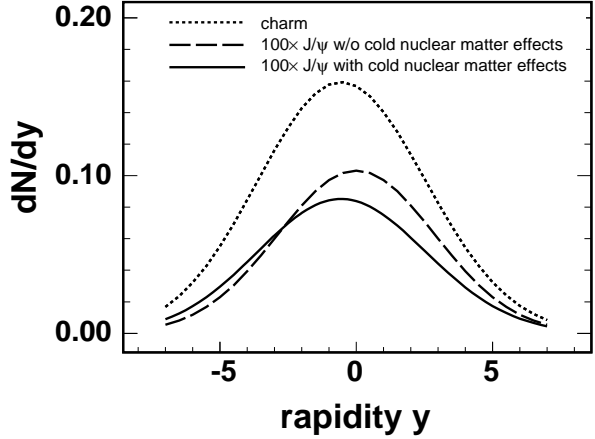}
    \caption{Rapidity distributions of initially produced $J/\psi$ with (solid line) and without (dashed line) cold nuclear matter effects as well as of charm quarks (dotted line) in the most central $10\%$ $p+\textrm{Pb}$ collisions at $\sqrt{s_{\textrm{NN}}}=5.02$ TeV. Positive and negative rapidities correspond to those of proton and Pb, respectively.}
    \label{fig_charm}
 \end{figure}
 
For the rapidity distribution, we take $d\sigma/dy=d\sigma/dy|_{y=0}\times e^{-y^2/(2\xi^2y_m^2)}$, and $D_y=D_0(1-y^2/y_m^2)=(n-2)\langle p_T^2\rangle_y$, where $y_m=\ln(\sqrt{s_{\textrm{NN}}}/m_{J/\psi})$ is the maximum rapidity that a $J/\psi$ can have due to the constraint from energy conservation, and $\xi=0.39$~\cite{Bossu:2011qe, Liu:2009wza}.  We then find $\langle p_T^2\rangle_{y=0}=9.5$ GeV$^2$.
From the average thickness function $T(\langle b\rangle_{0-10\%})=1.82$\ fm$^{-2}$ in a $p+\textrm{Pb}$ collision, where $b$ is the impact parameter of the collision, the rapidity distribution of initially produced $J/\psi$ can be determined and is shown by the dashed line in Fig.~\ref{fig_charm}. Including the cold nuclear matter effects by multiplying the above distribution with the nuclear modification factor estimated (see next section) from Refs.~\cite{Albacete:2013ei, Manceau:2013zta}, we obtain the solid line in Fig.~\ref{fig_charm}, which is used as the initial $J/\psi$ rapidity distribution for solving the transport equation. The cold nuclear matter effects, particularly the Cronin effect, can also modify the  transverse momentum distribution of initially produced $J/\psi$ in $p+\textrm{Pb}$ collisions.  From the average number of binary collisions per event $N_{\rm coll}=12.3$ in a central $p+\textrm{Pb}$ collision, which leads to a broadening of the transverse momentum $\Delta \langle p_T^2\rangle=4.2$\ GeV$^2$~\cite{Kang:2012am}, the mean-squared transverse momentum of initially produced $J/\psi$ is then increased to  $13.7$\ GeV$^2$.

As to the density distribution of initially produced $J/\psi$ in space, it is assumed to be proportional to the thickness of a uniform solid  sphere of radius $r=0.8$ fm as that for the proton.  

For the initial number of charm quarks, which is needed for calculating the $J/\psi$ regeneration rate, it is about $1.2$ pairs in a central $p+\textrm{Pb}$ collision, obtained by using  a total charm quark production cross section of  $6.8$ mb in a $p+p$ collision interpolated from the experimental data~\cite{Abelev:2012vra} and the average thickness function $T(\langle b\rangle_{0-10\%})=1.82$ fm$^{-2}$. The rapidity distribution of charm quarks  is assumed to be the same as that of the initially produced $J/\psi$ with cold nuclear matter effects, and this is shown by the dotted line in Fig.~\ref{fig_charm}. As to the  initial spatial distribution of charm quarks in the transverse plane, it is obtained  from the AMPT model and has a  qualitatively similar profile as  that of the initial energy density.

The above treatment for the ground state $J/\psi$ can be generalized to include its excited states $\chi_c$ and $\psi^\prime$ with different dissociaiton temperatures $T_D$ and cross sections~\cite{Wang:2002ck, Arleo:2001mp} as in Ref.~\cite{Liu:2012zw}. For the initial abundance of these excited states, they are determined from that of $J/\psi$ by using the empirically known feed-down contributions to $J/\psi$ production in  p+p collisions,  i.e., the proportion of direct $J/\psi$, feed-down from $\chi_c$ and that from $\psi^\prime$ is $6:3:1$~\cite{Zoccoli:2005yn}. They are further assumed to suffer the same cold nuclear matter effects as the $J/\psi$.
 
For the contribution to $J/\psi$ production from the decay of regenerated $\chi_c$ and 
$\psi^\prime$, they are  included with the branch ratios from Ref.~\cite{Beringer:1900zz}.

\section{Results}\label{se_results}

\begin{figure}[!hbt]
    \centering
    \includegraphics[width=0.48\textwidth]{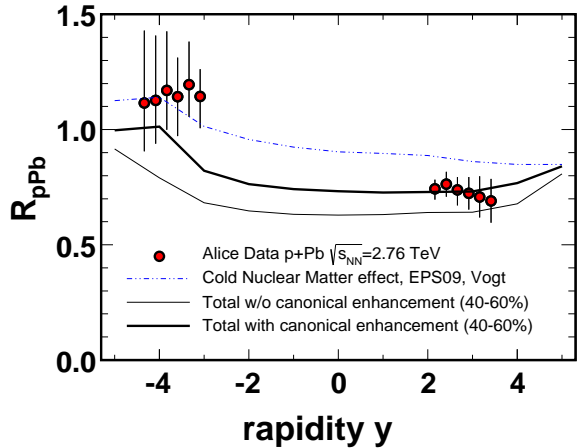}
    \caption{(Color online) Nuclear modification factor $R_{p{\rm Pb}}$\ of $J/\psi$ in the $40{\textrm -}60\%$ centrality bin of  $p+\textrm{Pb}$ collisions as a function of rapidity $y$ . The (blue)  dash-double-dotted line is for results with only cold nuclear matter effects~\cite{Albacete:2013ei}.  The (black) thin solid line and (black) thick solid line are the total $R_{p{\rm Pb}}$ without and with the canonical enhancement, respectively. Experimental data are for minimum bias collisions from the ALICE Collaboration~\cite{Manceau:2013zta}.}
    \label{fig_raa_mb}
\end{figure}

With the solution from the transport equation, observables on $J/\psi$ can  be calculated. We first  
show in Fig.~\ref{fig_raa_mb} the nuclear modification factor $R_{p{\rm Pb}}$ in the centrality bin of $40{\textrm -}60\%$, which has a  multiplicity of charged hadrons  very close to that in minimum-bias collisions~\cite{Albacete:2013ei} for which  $J/\psi$ production  has been measured by the ALICE Collaboration~\cite{Manceau:2013zta}. The nuclear modification factor $R_{p{\rm Pb}}$ only due to cold nuclear matter effects calculated in Refs.~\cite{Albacete:2013ei, Manceau:2013zta}, which is used to modify our initial $J/\psi$ distribution due to these effects as mentioned in the previous section, is denoted by the (blue) dash-double-dotted line\footnote{Values shown here  are averages of those in the upper and lower bands  shown in Ref.~\cite{Manceau:2013zta}.} and is seen to overestimate the results at forward rapidity. The (black) thin and thick solid lines are our results that include both initial cold nuclear matter effects and final hot  medium effects but without and with the canonical enhancement factor $C_{ce}=1+1/(dN_{c}^{\textrm{dir}}/dy)$ in Eq.~(\ref{eq_beta}), respectively.  Since the regeneration contribution is negligible without the canonical enhancement, the former result  reflects the contribution only from the initial production.   Comparing to the case including only the cold nuclear matter effects, $J/\psi$ production is seen to be suppressed in  the whole rapidity range of $-5<y<5$  as a result of the hot  medium effects. Furthermore, the suppression is stronger at backward rapidity due to the higher temperature of produced QGP, as already implied by the initial energy density distribution shown in Fig.~\ref{fig_epsilon}. By comparing the thin and thick solid lines, it can be seen that including the canonical enhancement for $J/\psi$ production increases the contribution from the regeneration, particularly  at backward rapidity where there is a  larger multiplicity of charm quarks relative to that of the $J/\psi$s from Glauber model as shown in Fig.~\ref{fig_charm}. With the inclusion of the hot medium effects, our results show that the experimental data can be better described.\footnote{Cold nuclear matter effects with energy loss~\cite{Arleo:2012rs} can also lead to a similar $R_{p\textrm{Pb}}$, indicating the large uncertainty in our current understanding of the cold nuclear matter effects.  We use the results from Ref.~\cite{Vogt:2004dh} for the cold nuclear effects because the same model describes reasonably well the rapidity distribution of $J/\psi$ in d+Au collisions at RHIC~\cite{Adare:2007gn}.  Also, our main interest in the present study is on the hot medium effects, rather the absolute value of the $R_{AA}$ of $J/\psi$.}

\begin{figure}[!hbt]
    \centering
    \includegraphics[width=0.48\textwidth]{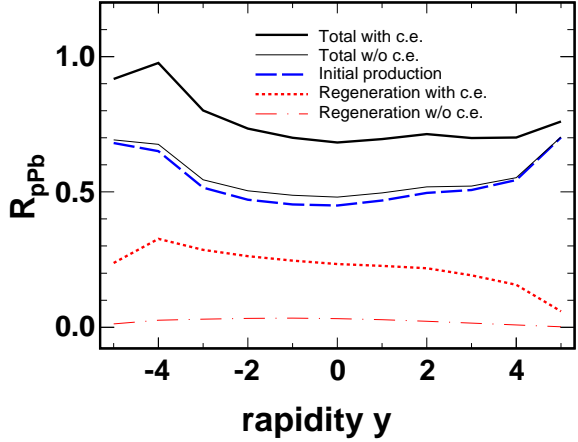}
    \caption{(Color online) Nuclear modification factor $R_{p{\rm Pb}}$ of $J/\psi$ in the most central 10\%  $p+\textrm{Pb}$ collisions as a function of rapidity $y$. The (blue)  dashed line is for initial production. The (red) dash-dotted line and (red) dotted line are the $R_{p{\rm Pb}}$ from regeneration without and with the canonical enhancement, respectively. The (black) thin solid line and (black) thick solid line are the total $R_{p{\rm Pb}}$ without and with the canonical enhancement, respectively.}
    \label{fig_raa}
\end{figure}

For the most central $10\%$  collisions, the initial energy density is roughly $3$ times larger. Because of the higher temperature in the produced QGP, stronger hot  medium effects are expected. To estimate the cold nuclear matter effects, we assume that they are  proportional to the thickness function $T(b)$.   With the average impact parameter $\langle b\rangle=3.3$ fm for the $0{\textrm -}10\%$ centrality collisions and $\langle b\rangle=5.7$ fm for the $40{\textrm -}60\%$ centrality collisions, the thickness function ratio in these two cases is  $C=T(\langle b\rangle_{0{\textrm -}10\%})/T(\langle b\rangle_{40{\textrm -}60\%})=1.85$. The nuclear modification factor of $J/\psi$ due to the cold nuclear matter effects is then estimated to be $R^{\rm cold}_{p\textrm{Pb}}(0{\textrm -}10\%)=1-C(1-R^{\rm cold}_{p\textrm{Pb}}(40{\textrm -}60\%))$. As shown by the dashed line in Fig.~\ref{fig_raa},  the initially produced $J/\psi$s suffer stronger suppression in these more central collisions. Although the cold nuclear matter effects increase the yield of $J/\psi$s at backward rapidity and reduce  that at forward rapidity, the suppression is also stronger at backward rapidity due to the higher temperature of the QGP. As a result, the initially produced $J/\psi$s do not differ much between forward and backward rapidities. Without the canonical enhancement, the  regeneration contribution in the most central collisions (dash-dotted line) is  still less than 0.1 and can be neglected as in semi-central collisions.  The regeneration contribution becomes, however, important after including the canonical enhancement effect, and the nuclear modification factor $R_{p{\rm Pb}}$  is again obviously larger at backward rapidity than  at forward rapidity as shown by the dotted line. The total $R_{p{\rm Pb}}$ including both cold nuclear matter and hot  medium effects in the case without and with the canonical enhancement are shown, respectively, by the thin and solid lines in Fig.~\ref{fig_raa}.

\begin{figure}[hbt]
     \centering
     \includegraphics[width=0.48\textwidth]{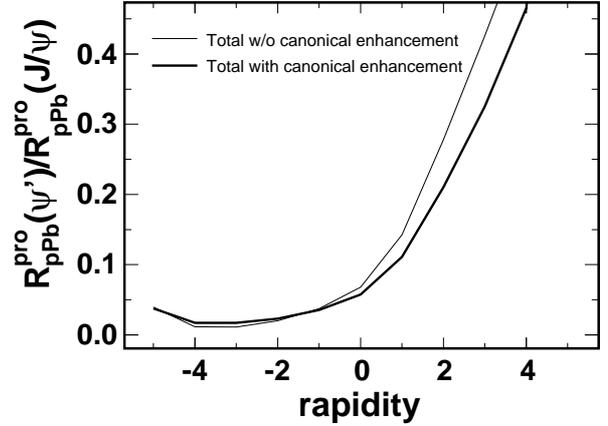}
     \caption{The ratio of the nuclear modification factor $R_{p\textrm{Pb}}$\ of prompt $\psi^{\prime}$\ to that of prompt $J/\psi$\ in the most central $10\%$  $p+\textrm{Pb}$ collisions at $\sqrt{s_{\textrm{NN}}}=5.02$ TeV as a function of rapidity $y$.}
     \label{fig_dr}
\end{figure}

 The above results indicate that the difference  between the $R_{p\textrm{Pb}}$s of $J/\psi$ with and without the hot  medium effects is less than $20\%$. Considering the uncertainties in theories and experiments, this difference in the $R_{p\textrm{Pb}}$ of $J/\psi$ is not sensitive enough to verify whether there  are hot  medium effects  in p+Pb collisions. Since  the hot medium effects on the ground state and the excited states of quarkonia,  are expected to be dramatically different according to the picture of sequential dissociation~\cite{Matsui:1986dk},  the double ratio $R_{p\textrm{Pb}}^{\textrm{pro}}(\psi^{\prime})/R_{p\textrm{Pb}}^{\textrm{pro}}(J/\psi)$ of prompt $\psi'$ to prompt $J/\psi$ would be a good probe to the hot medium effects. This is shown in Fig.~\ref{fig_dr}, where the double ratio is seen to be far less than one at the backward rapidity.This is because  the dissociation temperature $T_D$ of $\psi^{\prime}$ is only slightly higher than the critical temperature $T_c$, leading thus to its strong  suppression  at backward rapidity where the temperature of the fireball is high.  The value of the double ratio at the forward rapidity is, on other hand, much larger, since  the temperature of the fireball is lower and the hot  medium effects are less important. Experimental study of the double ratio $R_{p \textrm{Pb}}^{\textrm{pro}}(\psi^{\prime})/R_{p\textrm{Pb}}^{\textrm{pro}}(J/\psi)$  thus provides the possibility to verify the existence of the hot  medium effects in $p+\textrm{Pb}$ collisions at the LHC energy.

We note that although the contribution of regenerated $\psi^\prime$ to $J/\psi$ production is small, its
contribution to final $\psi^\prime$ is  large at backward rapidity where few initially produced $\psi^\prime$ survives. For example, the number of regenerated $\psi^\prime$ at $y=-2$ is about twice of the survived initially produced $\psi^\prime$.

\section{Conclusions}\label{se_conclusions}

We have studied the hot medium effects on $J/\psi$ production in $p+\textrm{Pb}$ collisions at $\sqrt{s_{\rm NN}}=5.02$ TeV in a transport approach based on a 2+1 dimensional ideal hydrodynamic model for the bulk dynamics and found that  the experimental data can be described with the inclusion of these effects.  Our results indicate that the suppression of $J/\psi$ production due to screening and thermal dissociation is more important than the contribution due to regeneration from the charm and anticharm quarks in the produced QGP,  which leads to a smaller $J/\psi$ nuclear modification factor $R_{p{\rm Pb}}$  than in the case of including only the cold nuclear matter effects.  We have also found that the  canonical enhancement effect due to smaller number of produced charm and anticharm quarks is important for the regeneration contribution since it would be negligible otherwise. We have further predicted the $R_{p{\rm Pb}}$ of $J/\psi$ for  the most central $10\%$  collisions and suggested the measurement of the double ratio $R_{p\textrm{Pb}}^{\textrm{pro}}(\psi^{\prime})/R_{p\textrm{Pb}}^{\textrm{pro}}(J/\psi)$\ in $p+\textrm{Pb}$ collisions, which  would be dramatically smaller than one if there exist the  hot  medium effects. 

\section*{Acknowledgements}
We thank Ralf Rapp for helpful comments. 
This work was supported by the U.S. National Science Foundation under Grant No. PHY-1068572, the US Department of Energy under Contract No. DE-FG02-10ER41682, and the Welch Foundation under Grant No. A-1358.

\end{document}